# Exploring the very high energy gamma-ray emission (E > 100 GeV) of the hard spectrum Fermi sources 1FGL J2001.1+4351 and B3-2247+381 with MAGIC

K. Berger
*Instituto de Astrofisica de Canarias and Universidad de La Laguna, E-38206 La Laguna, Spain*

G. Giavitto
*Institut de Física d'Altes Energies (IFAE), Edifici Cn, Universitat Autónoma de Barcelona (UAB), E-08193 Bellaterra, Barcelona, Spain*

E. Lindfors, L. Takalo
*Tuorla Observatory, University of Turku, FI-21500 Piikkiö, Finland*

D. Paneque
*SLAC National Accelerator Laboratory, Stanford University, Stanford, CA 94305, USA and Max-Planck-Institut für Physik, D-80805 München, Germany*

A. Stamerra
*Università di Siena, and INFN Pisa, I-53100 Siena, Italy*

On behalf of the MAGIC Collaboration

MAGIC, a stereoscopic cherenkov telescope array, sensitive to gamma-rays between 50 GeV and several tens of TeV, is ideally suited to observe promising Fermi LAT sources with a hard γ-ray spectrum. Here we discuss the discovery of very high energy γ-ray (VHE, E > 100 GeV) emission from the Fermi LAT sources 1FGL J2001.1+4351 and B3-2247+381 with MAGIC. 1FGL J2001.1+4351, recently identified as MG4 J200112+4352 (Bassani et al. 2010), is most likely a high peaked BL Lacertae object. The red shift of this source is still unknown, though the identification of the optical host galaxy suggests z < 0.2. MAGIC observations indicate short term variability, since the source showed a strong emission of 20% of the Crab Nebula flux above 90 GeV during the 16th of July 2010 and none of the other observation nights yielded a detection. B3-2247+381 is classified as a BL Lac object at z = 0.1187 (Veron-Cetty & Veron catalogue of known AGN). In July 2010 it showed increased optical activity in the Tuorla blazar monitoring program, which subsequently activated target of opportunity observations by MAGIC. Within 18 hours of observation time extended over 13 days between September and October 2010, a strong signal was found above an energy threshold of 150 GeV. The flux (4% of the Crab Nebula) is consistent with being constant over the entire observation campaign. We compute the light curves, model the spectral energy distributions of these new very high energy γ-ray emitters and discuss the physical properties of the VHE γ-ray emission region.

## 1. INTRODUCTION

In recent years the window to the very high energy (VHE, E > 100 GeV) γ-ray sky has been opened wide: up to now 125 sources of VHE γ-rays have been discovered[1]. Since however the most sensitive instruments are designed for pointed observations with a typical field of view in the order of a few degrees, full sky surveys are practically unfeasible. It is thus common practice to use already existing catalogues in the optical, X-ray or high energy range to determine potential targets for pointed observations with VHE γ-ray telescopes.

MAGIC, a system of two 17 m diameter Imaging Atmospheric Cherenkov Telescopes (IACT) of the latest generation, is sensitive to γ-rays between 50 GeV up to several tens of TeV. A summary of its current performance and software developments can be found in [1] and [2], respectively. In short, a good sensitivity of 0.76% (above ~300 GeV) of the Crab Nebula flux with an angular resolution better than 0.07° and an energy resolution as good as 16% have been achieved. In this proceeding we discuss the discovery of two new VHE γ-ray emitters by MAGIC: B3 2247+381 and 1FGL J2001.1+4351.

## 2. B3-2247+381

A redshift of z = 0.119 [3], an X-ray flux > 2 μJy [4] and the classification as high frequency peaked BL Lac object [5] made B3-2247+381 a promising target for VHE γ-ray observations. The source was also included in the list of potential TeV sources released to the IACT experiments by the FERMI - Large Area Telescope (FERMI-LAT) Collaboration in October 2009 [6]. Observations with the single MAGIC-I telescope for 8.3 hours in 2006 did not result in a detection and thus an upper limit of 5.2% of the Crab Nebula flux above 140 GeV was derived [7].

In September 2010, B3-2247+381 showed a significant increase of its emission in the R-band[2], which triggered observations with MAGIC and Swift. MAGIC observed the source for a total of 18.3 h and detected VHE γ-ray emission at a significance level of 5.6 σ with an estimated flux of 2% of the Crab Nebula flux above 150 GeV ([8], Figure 1). The measured flux is consistent with the upper limit from the 2006 observations. It is thus unclear if the source was in a high VHE γ-ray state. However the X-ray light curve shows a significant enhancement of the flux during our observations as well as variability (see Figures 2 and 3).

---

[1] TevCat: http://tevcat.uchicago.edu/

[2] http://users.utu.fi/kani/1m/index.html

The differential energy spectrum is well described by a simple power-law. The photon index was found to be $-3.2 \pm 0.5_{stat} \pm 0.5_{sys}$ with a flux normalization $f_0$ at 300 GeV of $(1.4 \pm 0.3_{stat} \pm 0.2_{sys})\, 10^{-11}$ ph cm$^{-2}$ s$^{-1}$ TeV$^{-1}$. Taking into account the attenuation due to pair production with the extragalactic background light (EBL), the spectrum was found to be compatible with a power law with a photon index $\gamma = -2.7 \pm 0.5_{stat} \pm 0.5_{sys}$ and a flux at 300 GeV $f_0 = (2.0 \pm 0.3_{stat} \pm 0.3_{sys})\, 10^{-11}$ ph cm$^{-2}$ s$^{-1}$ TeV$^{-1}$. Two different EBL models were used ([9] and [10]), and the obtained results were found to be in good agreement with each other, well within the statistical uncertainties. The finally obtained VHE γ-ray spectrum is shown in Figure 4. The black data points refer to the measured spectrum, while the grey dashed points have been corrected for the attenuation of the EBL. The thick black and dashed grey lines are power-law fits to the respective data points. The dashed band corresponds to the statistical error of the fit to the measured spectrum, while the white band surrounding it is the sum of the statistical and systematic errors of the fit.

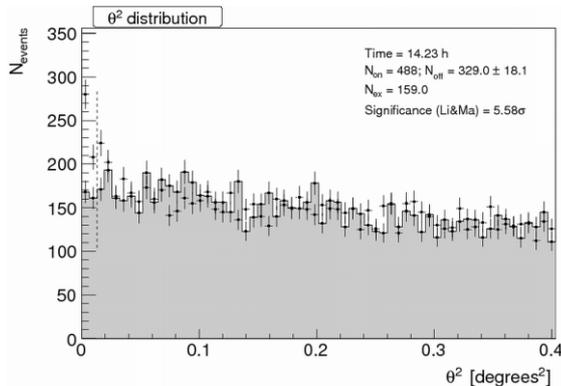

Figure 1: Distribution of the squared angular distance ($\theta^2$) for the on-source counts in the direction of B3-2247+381 (black points with error bars) and the normalized off-source events (gray histogram).

In Figure 5 we show the spectral energy distribution of the source during the MAGIC observations, together with simultaneous Swift and optical data, and other non-simultaneous data. The Swift observations show that the source was in a high state also in X-rays. In the Fermi energy range the source is very weak, which limits the capability of detecting statistically significant flux-variability on time scales of a few months. The synchrotron component of the SED is showing a significantly larger emission in the high state, while the inverse Compton component is consistent with only minor changes. We must however note that the weak detection in the HE and VHE γ-ray band significantly limits the determination of the inverse Compton component.

We reproduce the SED with a one-zone synchrotron-self Compton (SSC) model (see [11] for a description). In brief, the emission region is assumed to be spherical, with a radius $R$, filled with a tangled magnetic field of intensity $B$. The relativistic electrons follow a smoothed broken powerlaw energy distribution specified by the limits $\gamma_{min}$, $\gamma_{max}$ and the break at $\gamma_b$ as well as the slopes $n_1$ and $n_2$ before and after the break, respectively. Relativistic effects are taken into account by the Doppler factor. The used input model parameters are shown in Table 1.

The comparison with parameters derived for BL Lac objects (e.g. [12]) reveals that the parameters used for B3-2247+381 are close to the typical values. As for other sources, the somewhat larger (lower) value of the Doppler factor (the magnetic field intensity B) with respect to "standard" values is mainly due to the relatively large separation between the synchrotron and IC peaks.

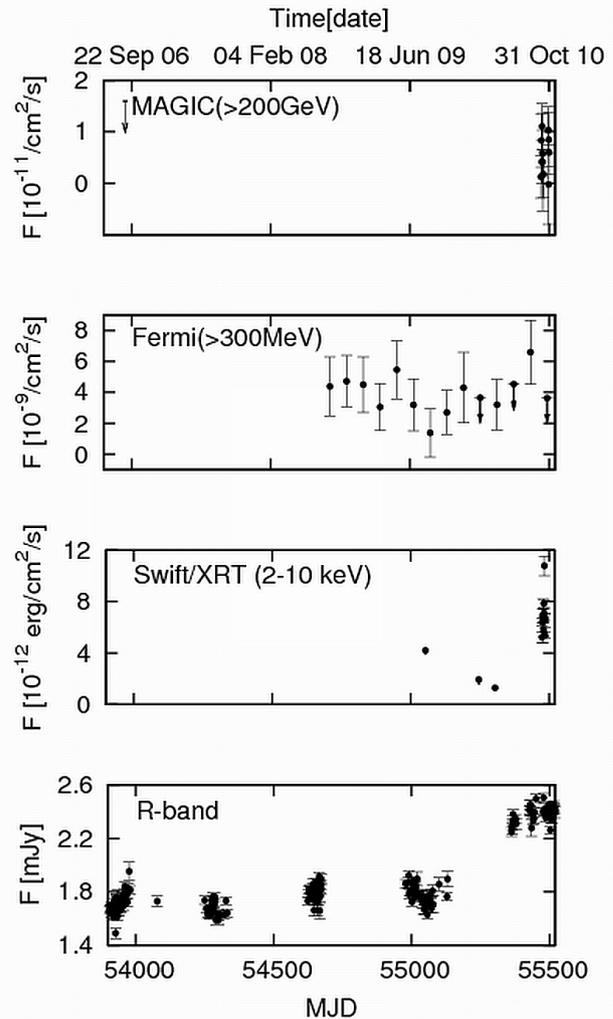

Figure 2: Long term light curves of B3-2247+381 in VHE γ-rays, Fermi-LAT HE γ-rays (two months time intervals), Swift X-rays and optical KVA R-band.

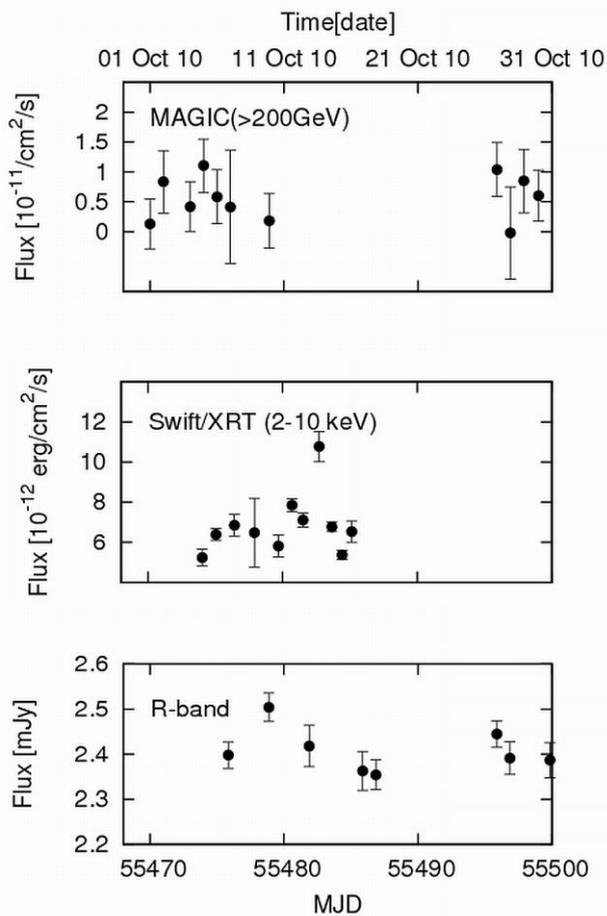

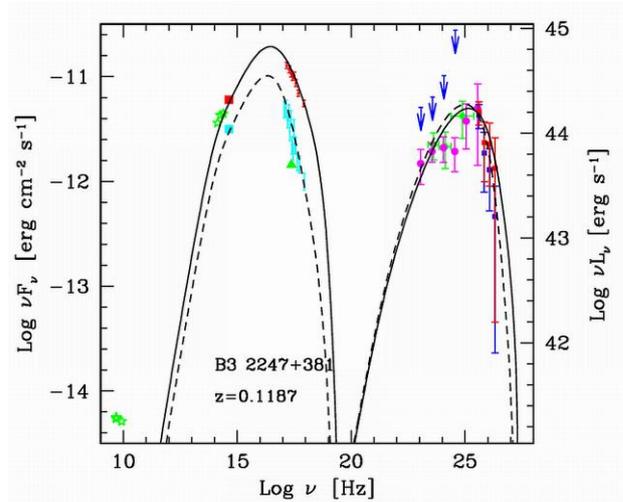

Figure 5: Spectral energy distribution of B3-2247+381 (red: EBL corrected MAGIC spectral points). The green crosses are 1FGL Fermi data points (Abdo et al. 2010), while the pink points represent the Fermi analysis of this work (2.5 years of data). Blue arrows show the 95% confidence upper limits computed from Fermi-LAT data for the time interval of the MAGIC observation. Low (high) state Swift data were taken on April 18th 2010 (October 5-16, 2010). The host galaxy contribution has been subtracted from the KVA R-band data (red and light blue squares), following Nilsson et al. (2007). The data have been corrected for galactic absorption. Green and light blue points represent non-simultaneous low state data. The solid line is our SSC-model fit to the high state observations; the dotted line is a fit to the low state observations.

Figure 3: Same as Fig. 2, but zoomed into the time interval of the MAGIC observations in September-October 2010.

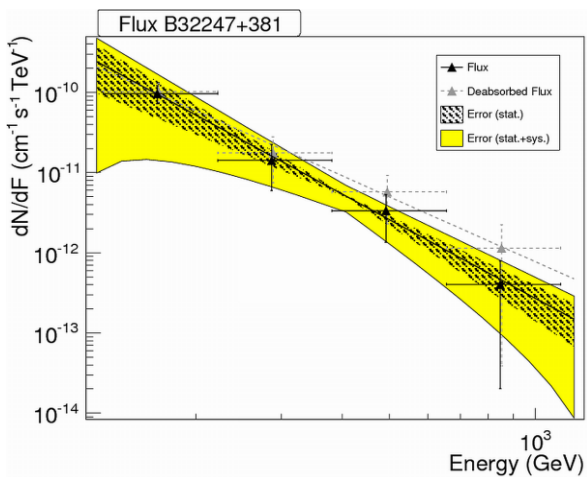

Figure 4: The unfolded differential energy spectrum of B3-2247+381 observed by MAGIC. For more explanations see text.

| Flux State | $\gamma_{min}$ | $\gamma_b$ | $\gamma_{max}$ | $n_1$ | $n_2$ | B G | K $cm^{-3}$ | $\delta$ | R cm |
|---|---|---|---|---|---|---|---|---|---|
| High | $3 \cdot 10^3$ | $7.1 \cdot 10^4$ | $6 \cdot 10^5$ | 2.0 | 4.35 | 0.06 | $2.5 \cdot 10^3$ | 35 | $8 \cdot 10^{15}$ |
| Low | $3 \cdot 10^3$ | $6.8 \cdot 10^4$ | $5 \cdot 10^5$ | 2.0 | 5.35 | 0.08 | $1.15 \cdot 10^4$ | 30 | $4 \cdot 10^{15}$ |

Table 1: Input parameters for the high and low states of the SSC-model shown as solid and dashed lines in Figure 5. For more explanations see text.

## 3.  1FGL J2001.1+4351

Following an agreement with the IACT groups (MAGIC, VERITAS and H.E.S.S.) the FERMI-LAT Collaboration provided a list of sources flagged as good VHE-candidates, which included 1FGL 2001.1+4351 (RA=300.304, Dec=43.886) as one of the most prominent candidates [6]. Remarkably it was at first detected only above 1 GeV [13]. The unknown identification of the object triggered a study by Bassani et al. [14] in the optical, radio and X-ray band. The most likely counterpart, MG4 J200112+4352 (a bright flat spectrum radio source), was subsequently classified as a high frequency peaked BL Lacertae object. The source was found to be variable both in the X-ray as well as the optical band. The redshift of this source is still unknown (identification of the optical host galaxy suggests z<0.2).

MAGIC observed 1FGL J2001.1+4351 from July until September 2010. Only one night, July 16$^{th}$, showed a significant excess of 7.6 σ (pre-trial, Figure 6, [15]). This corresponds to 22% of the Crab Nebula flux above 90 GeV. The preliminary analysis of *Swift* data obtained simultaneously with MAGIC indicates an enhancement in the X-ray flux by a factor of three with respect to the previous days, indicating a positive correlation of the VHE and X-ray flux for this source.

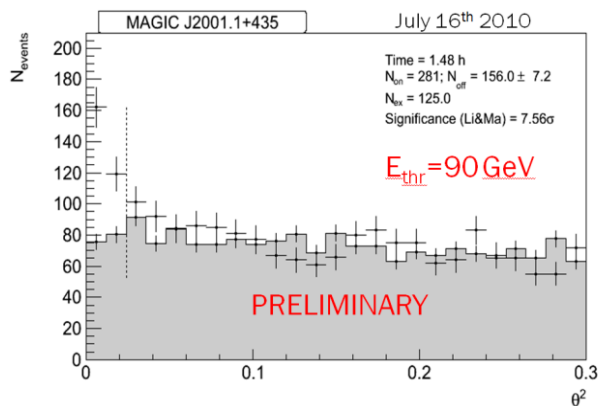

Figure 6: Squared angular distribution with respect to the position of 1FGL J2001.1+4351 (Observation details in the inlay). The energy threshold of this preliminary analysis is ~90 GeV.

## 4.  CONCLUSIONS

The two sources discovered were detected during high X-ray states (B3-2247 additionally showed enhanced optical emission). It can thus be concluded that optical/X-ray triggers and a pre-selection based on the FERMI γ-ray catalogue are successfully enhancing the detection probability of VHE γ-ray sources.

The final results (including the spectral energy distributions) will be discussed in forthcoming journal publications.

## Acknowledgments

We would like to thank the Instituto de Astrofísica de Canarias for the excellent working conditions at the Observatorio del Roque de los Muchachos in La Palma. The support of the German BMBF and MPG, the Italian INFN, the Swiss National Fund SNF, and the Spanish MICINN is gratefully acknowledged. This work was also supported by the Marie Curie program, by the CPAN CSD2007-00042 and MultiDark CSD2009-00064 projects of the Spanish Consolider-Ingenio 2010 programme, by grant DO02-353 of the Bulgarian NSF, by grant 127740 of the Academy of Finland, by the YIP of the Helmholtz Gemeinschaft, by the DFG Cluster of Excellence "Origin and Structure of the Universe", by the DFG Collaborative Research Centers SFB823/C4 and SFB876/C3, and by the Polish MNiSzW grant 745/N-HESSMAGIC/2010/0.